# Networks Become Navigable as Nodes Move and Forget

Augustin Chaintreau[*]    Pierre Fraigniaud[†]

Emmanuelle Lebhar[‡]

**Abstract**

We propose a dynamical process for network evolution, aiming at explaining the emergence of the small world phenomenon, i.e., the statistical observation that any pair of individuals are linked by a short chain of acquaintances computable by a simple decentralized routing algorithm, known as greedy routing. Previously proposed dynamical processes enabled to demonstrate experimentally (by simulations) that the small world phenomenon can emerge from local dynamics. However, the analysis of greedy routing using the probability distributions arising from these dynamics is quite complex because of mutual dependencies. In contrast, our process enables complete formal analysis. It is based on the combination of two simple processes: a random walk process, and an harmonic forgetting process. Both processes reflect natural behaviors of the individuals, viewed as nodes in the network of inter-individual acquaintances. We prove that, in $k$-dimensional lattices, the combination of these two processes generates long-range links mutually independently distributed as a $k$-harmonic distribution. We analyze the performances of greedy routing at the stationary regime of our process, and prove that the expected number of steps for routing from any source to any target in any multidimensional lattice is a polylogarithmic function of the distance between the two nodes in the lattice. Up to our knowledge, these results are the first formal proof that navigability in small worlds can emerge from a dynamical process for network evolution. Our dynamical process can find practical applications to the design of spatial gossip and resource location protocols.

**Keywords:** Small world phenomenon, dynamical process, routing, spatial gossip, resource location, random walks.

[*]Thomson, Paris. Email: *Augustin.Chaintreau@thomson.net*.

[†]CNRS and University Paris 7. Email: *Pierre.Fraigniaud@liafa.jussieu.fr*. Additional supports from the ANR projects ALADDIN and ALPAGE, and from the COST Action 295 DYNAMO.

[‡]CNRS and University Paris 7. Email: *Emmanuelle.Lebhar@liafa.jussieu.fr*. Additional supports from the ANR project ALADDIN, and from the COST Action 295 DYNAMO.

# 1 Introduction

Models relating geography and social-network friendship enable a good understanding of the small world phenomenon, a.k.a., six degrees of separation between individuals [11, 29]. In these models, the probability of befriending a particular person is assumed to be inversely proportional to the number of closer people, fitting with what was observed experimentally (cf. [28]). Under this assumption, it was proved that, using ad hoc probability distributions, many classes of graphs are navigable, that is, a simple decentralized routing procedure enables efficient routing from any source to any target. (By efficient, we mean, as it is standard in this framework, that routing from any source $s$ to any target $t$ takes a polylogarithmic expected number of steps). For instance, such a navigability property is satisfied in multi-dimensional meshes [24], in graphs of bounded ball growth [13], and more generally in graphs of bounded doubling dimension [34]. In all these cases, a graph $G$, that may not only represent geography but also other proximity measures like professional activities, religious beliefs, etc., is enhanced with additional links chosen at random. More precisely, every node is given some *long-range links* pointing at other nodes in the graph. For each long-range link added at a node $u$, the probability that the head of this link is $v$ is inversely proportional to the size of the ball of radius $\text{dist}_G(u, v)$ centered at $u$ in $G$, hence depending on the density of $G$ around $u$. This setting applies to weighted graphs too [26], and to infinite graphs as well [13]. For instance, in the $k$-dimensional lattice $\mathbb{Z}^k$, the probability that $u$ has a long-range link pointing at $v$ is essentially proportional to $1/d^k$ where $d$ is the distance between $u$ and $v$ in the lattice. This setting of the long-range links enables greedy routing[1] to perform in polylogarithmic expected number of steps (as a function of the distance in the lattice between the source and the target).

## 1.1 Navigability as an emerging property

In [25] (Problem 7), Jon Kleinberg asks about "what kinds of growth processes or selective pressures might exist to cause networks to become more efficiently searchable". Many attempts have been made to explain how the density-based distribution of the long-range links can emerge with time from the evolution of a network. Inspired by the world wide web or by P2P file-sharing systems, all the models we are aware of have considered the augmentation process (or rewiring) of a static graph used by its nodes for searching information. Our work uses a different approach, starting from the following observations. One the one hand, anyone of us can call or email any person in the world. On the other hand, to do so, it is frequently the case that we have met this person before. We thus start from the assumption that long-range connections are between remote people who have met once in the past. In other words, long-range links are emerging from nodes mobility, that we model by random walks in this paper. Another observation is that people forget some of their former acquaintances along with time. This forgetting mechanism represents the well understood fact that one cannot maintain close relationships with an explosive number of people. Thus we couple the random walk process with a forgetting process, and prove that this idealistic setting is sufficient to insure polylogarithmic navigability with simply one long-range connection per node.

---

[1]Greedy routing [24] aims at modeling the routing strategy performed by the individuals in Milgram experiment. In a graph $G$ enhanced with long-range links, a node $u$ handling a message of destination $t$ selects among all its neighbors, including its long-range contact(s), the one that is the closest to the target $t$ according to the distance in the base graph $G$, and forwards the message to that node.

## 1.2 Rewiring processes

Clauset and Moore [9] proposed the following rewiring process for the multidimensional lattice, inspired by the actions of surfers on the web. While routing from a source $s$ to a target $t$, if the target is not reached after $\tau$ steps, then the long-range link of $s$ is rewired to point at the current node $x$. The threshold $\tau$ is set based on the distance (in the lattice) between $s$ and $t$, and on the expected time of greedy routing from $s$ to $t$ when the $k$-dimensional lattice is augmented using the $k$-harmonic distribution [24]. The simulation results presented in [9] show that the distribution $f$ of the link lengths converges to the power law $h(d) \propto 1/d^k$. Sandberg and Clarke [32] proposed a different rewiring process, based on Freenet feedback mechanisms [8]. This iterative process selects, at each phase, two nodes $s$ and $t$ uniformly at random, and constructs the greedy path $s = x_0, x_1, \ldots, x_{k-1}, x_k = t$ from $s$ to $t$. For every $i \in \{0, 1, \ldots, k\}$, the long-range link of $x_i$ is rewired with probability $p$, to point at $t$. The $k+1$ decisions (rewiring or not) are taken mutually independently. This process is analyzed in [31]. It is proved that, under some hypotheses, the process converges. Moreover, the stationary distribution $f$ of the link lengths can be fully characterized. In the $k$-dimensional lattice, it is close to the power law $h(d) \propto 1/d^k$ for an appropriate $p$, and simulations show that greedy routing in rings and meshes enhanced using the stationary distribution $f$ performs as efficiently as when these networks are enhanced using the 1- and 2-harmonic distributions, respectively.

For both [9] and [32], the complete formal analysis of the process remains open (even the formal characterization of the stationary distribution of the processes described in [9] remains open). The difficulty of the analysis is due to the dependencies between the long-range links generated by the processes. In particular, the computation of the greedy routing performances is a challenge when the long-rank links are not mutually independent. So, building further theory upon these two models looks quite difficult.

In this paper, we propose a dynamical network model based on the com-bination of two simple processes: a random walk process, and a harmonic forgetting process.We prove that the combination of these two processes generates long-range links mutually independently distributed as a distribution that resembles the density-based distribution, and from which navigability provably emerges.

### 1.3  Sketch of our network evolution process

In our network evolution process, called move-and-forget, or M&F for short, individuals are modeled by tokens moving from node to node in the $k$-dimensional lattice $\mathbb{Z}^k$, for some fixed integer $k \geq 1$ (the dimension of the lattice may be related to the number of proximity criteria used by the individuals for routing). Initially, each node is occupied by exactly one token. These tokens are moved mutually independently during the execution of the dynamical process, according to a random walk.

Tokens are attached to the heads of the long-range links, whose tails are the nodes from where the tokens initially started their random walks. Using the analogy of individuals moving in the geographical world, each long-range link indicates an acquaintance between an individual located at a fixed geographical point (where the token initially stood) and some individual located at some geographical coordinates (where the token currently stands).

The random walk process is coupled with another dynamic: nodes may forget their contacts through their long-range links. The motivation for our forgetting process is that individuals may loose contact with former good friends, but they meet new people among which some may become close friends. Since older acquaintances indicate stronger relationships, we assume that they have less probability to be forgotten than recent ones. More precisely, a long-range link of age $a$, that is a long-range link that survived $a$ steps of the forgetting process, is forgotten with probability $\phi(a) \propto 1/a$. When a long-rang link is forgotten by a node, it is rewired to point at this node (hence creating a self-loop). The token at the head of the forgotten link is removed, and a new token is launched at the node. (A new local relationship replaces an old remote relationship).

Note that M&F is defined independently from the dimension $k$ of the lattice: tokens execute random walks, and they are forgotten with a probability that depends only of their ages.

### 1.4  Our results

We prove that, for any fixed integer $k \geq 1$, the M&F rewiring process sketched above converges in the $k$-dimensional lattice to a distribution $f$ of the link lengths that resembles the $k$-harmonic distribution. Precisely, we prove that there exists $d_0 \geq 0$ and two positive constants $c$ and $c'$, such that, for any $\mathbf{d} = (d_1, \ldots, d_k) \in \mathbb{Z}^k$ with $|d_i| \geq d_0$ for all $i \in \{1, \ldots, k\}$, we

|  | Convergence | Navigability |
| --- | --- | --- |
| A. Clauset and C. Moore (2003) | Simulations | Simulations |
| O. Sandberg and I. Clarke (2007) | **Proof** | Simulations |
| Move-and-forget (M&F) | **Proof** | **Proof** |

Table 1: Properties of known network evolution processes compared to M&F

have
$$\frac{c}{\|\mathbf{d}\|^k \cdot \ln^{1+\epsilon} \|\mathbf{d}\|} \leq f(\mathbf{d}) \leq \frac{c' \ln^{k/2} \|\mathbf{d}\|}{\|\mathbf{d}\|^k \cdot \ln^{1+\epsilon} \|\mathbf{d}\|}$$

where $\epsilon > 0$ is a fixed (arbitrary small) parameter of M&F, and $\|\cdot\|$ denotes the $\ell_\infty$ norm.

Moreover, M&F guarantees the mutual independence of the long-range links. As a consequence, the performances of greedy routing in the lattice enhanced using the distribution $f$ can be analyzed formally. We prove that the expected number of steps of greedy routing from any source $s$ to any target $t$ at distance $d$ in the $k$-dimensional lattice satisfies

$$\mathbb{E}[X_{s,t}] \leq O(\ln^{2+\epsilon} d).$$

Therefore, greedy routing performs polylogarithmically as a function of the distance between the source and the target. In particular, the performances of greedy routing are essentially the same as the ones obtained by Kleinberg [24] using the ad hoc $k$-harmonic distribution [24].

Up to our knowledge, these results are the first formal proof that navigability in small worlds can emerge from a dynamical process for network evolution (see Table 1). Moreover, M&F is simple (by just coupling two simple dynamics), naturally distributed (each node takes care of just its token), robust (the loss of one token simply requires to launch a new token), and scalable (by direct adaptations of the infinite lattice setting to square toroidal meshes of arbitrary sizes).

Last but not least, M&F can find practical applications, including the design of distributed spatial gossip and resource location protocols.

## 1.5 Related works

The search for a network evolution process that could explain the emergence of the small world phenomenon in social networks started with the pioneering work of Watts and Strogatz [35] who proposed a rewiring process in the cycle, generating networks possessing small diameter and large clustering coefficient, simultaneously. Adding random matchings to cycles, as in [5], yields graphs with small diameter, but non necessarily with small clustering coefficient. As far as navigability is concerned, these networks do not support efficient decentralized routing mechanisms [24]. Albert and Barabási [2] produced a thorough investigation of the preferential attachment model [33]. Although the preferential attachment model enables the design of efficient search procedures under specific circumstances (see [16] and the references therein), the recent lower bounds in [12] show that polylogarithmic routing cannot be achieved in general in networks generated according to this model. Recently, Liben-Nowell and Kleinberg [27] tried to infer which interactions in social networks are likely to occur in the near future from the observation of the existing ones, but navigability is not of their concern. Actually, as far as we know, the only network evolution models from which polylogarithmic routing emerges are the aforementioned ones [9, 32], which we already discussed.

Following up the seminal work of Kleinberg [24], a large literature has been dedicated to the analysis of greedy routing in graphs enhanced by long-range links set according to various kinds of probability distributions (see, e.g., [1, 13, 14, 15, 34]). These papers proved that several large classes of graphs can be enhanced by long-range links so that greedy routing performs in polylogarithmic expected number of steps. A lower bound of $\Omega(n^{1/\sqrt{\log n}})$ expected number of steps for greedy routing in arbitrary graphs has been proved in [18], and an upper bound of $O(n^{1/3})$ has been proved in [17]. Lower bounds for the cycle can be found in [3, 4, 19].

## 2 The Move-and-Forget (M&F) Rewiring Process

### 2.1 Process description

#### 2.1.1 Random walks

Let $k \geq 1$ be an integer. The rewiring process move-and-forget (M&F for short) assumes that each node in the $k$-dimensional lattice $\mathbb{Z}^k$ is initially occupied by exactly one token. These tokens move mutually independently according to random walks. That is, each token is given a set of $k$ fair coins $c_i$, $i = 1, \ldots, k$. At each step of its walk, each token flips its $k$ coins, and moves in the $i$th dimension of the lattice in the positive direction if $c_i$ is head, and in the negative direction if it is tail. More precisely, let $X(t) \in \mathbb{Z}^k$ denotes the position of a token in the lattice after $t$ steps of M&F, assuming that the token initially started at node $(u_1, \ldots, u_k) \in \mathbb{Z}^k$. We have $X(0) = (u_1, \ldots, u_k)$, and, for $t \geq 1$, $X(t) = (X_1(t), \ldots, X_k(t))$ satisfies

$$X_i(t) = \begin{cases} X_i(t-1) + 1 & \text{with probability } 1/2; \\ X_i(t-1) - 1 & \text{with probability } 1/2. \end{cases} \quad (1)$$

#### 2.1.2 Setting of the long-range links

Tokens are attached to the heads of the long-range links, whose tails are the nodes from where the tokens initially started their random walks (see Figure 1(a)). The head of a long-range link is called the long-range *contact* of the tail of this link. Hence the long-range contact of a node $u$ is the node $v$ currently occupied by the token launched by node $u$.

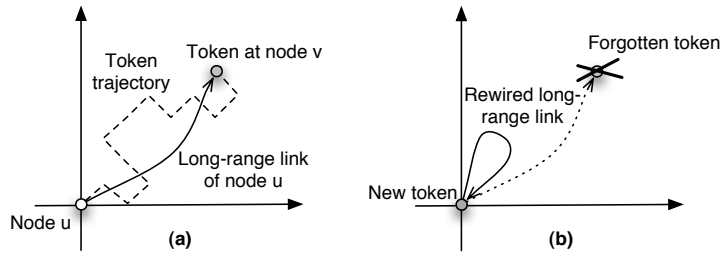

Figure 1: Dynamic of the long-range links in M&F.

### 2.1.3 Forgetting process

Nodes may forget their contacts through their long-range links. More precisely, a long-range link of age $a \geq 0$, that is a long-range link that survived $a$ steps of the forgetting process, is forgotten with probability $\phi(a)$. When a long-range link is forgotten by a node, it is rewired to point at this node (see Figure 1(b)). The token at the head of the forgotten link is removed, and a new token is launched at the node. This new token starts another random walk in $\mathbb{Z}^k$. Hence, if $A(t) \in \mathbb{N}$ denotes the age of the long-range link of some node $u$, that is the number of steps between time $t$ and the last time this link was rewired during the execution of M&F, and if $C(t)$ denotes the long-range contact of node $u$ at step $t$, then we have $C(t) = X(A(t))$.

The forgetting function $\phi$ has a huge impact on the distribution of the long-range link lengths. In this paper, we will consider $\phi(a) \propto 1/a$. The precise setting of $\phi$ will appear more complex for technical reasons only[2] (series convergence for infinite lattices, normalization, etc.). In fact, its behavior essentially reflects a decreasing of the forgetting probability that is inversely proportional to the age of the relationships. The precise setting of $\phi$ is described in the next section which explains the connections between the random walk $X$, the forgetting function $\phi$, and the distribution $f$ of the long-range link lengths.

## 2.2 Setting of the forgetting function

We first prove that the age of the long-range link resulting from the execution of M&F at a node has a stationary distribution (the proof of this lemma can be found in Appendix A).

---

[2]For instance, one needs $\sum_{a \geq 0} \phi(a)$ to diverge since otherwise the Markov chain $A(t)$ would be transient, and links could survive infinitely with positive probability. However, on the one hand, just setting $\phi(a) = 1/a$ would make $A(t)$ recurrent null (and thus for any $a$ we would have $\Pr\{A(t) = a\}$ converging to 0 as $t$ goes to infinity), but, on the other hand, setting $\phi(a) = 1/a^\alpha$ with $\alpha < 1$ would not yield navigability.

**Lemma 1** *For any function $\phi$ in $[0,1)$ such that the series of general term $\Pi_{i=1}^{j}(1-\phi(i))$ is finite, $(A(t))_{t\geq 0}$ is a Markov chain which is irreducible, aperiodic, and positive recurrent, with stationary probability distribution $\pi$ where*
$$\pi(a) = \frac{\Pi_{i=1}^{a}(1-\phi(i))}{\sum_{j\geq 0}\Pi_{i=1}^{j}(1-\phi(i))},$$
*for all $a \geq 0$.*

**Definition 1** *We define the forgetting probability $\phi$ as the following function:*
$$\phi(a) = \begin{cases} 0 & \text{if } a = 0, 1, \text{ or } 2; \\ 1 - \frac{a-1}{a}\left(\frac{\ln(a-1)}{\ln a}\right)^{1+\epsilon} & \text{if } a \geq 3; \end{cases} \quad (2)$$
*where $\epsilon > 0$ is arbitrary small.*

Note that $\phi(a) = \frac{1}{a} + o\left(\frac{1}{a}\right)$. Indeed,
$$\left(\frac{\ln(a-1)}{\ln a}\right)^{1+\epsilon} = \left(1 + \frac{\ln(1-1/a)}{\ln a}\right)^{1+\epsilon} = 1 - \frac{1+\epsilon}{a\ln a} + o\left(\frac{1}{a\ln a}\right)$$

If $\phi$ is defined according to Eq. (2), then Lemma 1 enables to give a close formula for $\pi$ (the proof of this lemma can be found in Appendix B).

**Lemma 2** *If $\phi$ is defined according to Eq. 2, then there exists a constant $c > 0$ such that $\pi(0) = \pi(1) = \pi(2) = c$ and for any $a \geq 3$,*
$$\pi(a) = \frac{c}{a \ln^{1+\epsilon} a}.$$

Finally, the relationship between the stationary distribution of the long-range link ages and the stationary distribution of the long-range link lengths is made explicit in the following lemma (see proof in Appendix C).

**Lemma 3** *The distribution of the long-range links converges to the distribution $f$ satisfying, for any $\mathbf{d} \in \mathbb{Z}^k$,*
$$f(\mathbf{d}) = \sum_{a\geq 0} \pi(a) \cdot \Pr\{X(a) = \mathbf{d}\}.$$

## 3  Analysis of the dynamical process M&F

In this section, we analyze the stationary distribution of the long-range link lengths in the $k$-dimensional lattice, and prove that this distribution resembles the $k$-harmonic distribution.

**Theorem 1** *There exist $d_0 \geq 0$ and two positive constants $c$ and $c'$ such that, for any $\mathbf{d} = (d_1, \ldots, d_k) \in \mathbb{Z}^k$ with $|d_i| \geq d_0$ for all $i \in \{1, \ldots, k\}$, we have*

$$\frac{c}{\| \mathbf{d} \|^k \cdot \ln^{1+\epsilon} \| \mathbf{d} \|} \leq f(\mathbf{d}) \leq \frac{c' \ \ln^{k/2} \| \mathbf{d} \|}{\| \mathbf{d} \|^k \cdot \ln^{1+\epsilon} \| \mathbf{d} \|}$$

*where $\epsilon > 0$ is the fixed parameter of* M&F*, and $\| \cdot \|$ denotes the $\ell_\infty$ norm.*

To prove the theorem, we first prove that, for large distances $d$, a random walk of age $a$ cannot be of length $d$ unless $a \geq \Omega(d^2)$. More precisely, we establish an exponentially small upper bound for the probability for a long-range link to be of length $d$ at age $a = o(d^2)$. Second, we prove that if the age $a$ is sufficiently large, then the chance for a random walk to reach a given distance $d$ at age $a$ is proportional to $\frac{1}{\sqrt{a}}$. Summing this probability over all values of $a$ larger than $d^2$ allows us to conclude that the transform of the age distribution $\pi$ described in Lemma 3 is approaching the $k$-harmonic distribution.

Let us establish some basic properties satisfied by random walks in dimension 1. We will extensively use the following Chernoff bound. Let $T$ be a sum of Bernouilli variables, with expectation $\mu$. Then [30]:

$$\Pr\{|T - \mu| > t\} \leq 2 \, \exp(-\frac{t^2}{4\mu}) \ \ \text{for any} \ \ t \leq \mu. \tag{3}$$

The following lemma specifies what must be the minimum order of magnitude for $a$ in order to contribute significantly to the sum defining $f$ in Lemma 3.

**Lemma 4** *Let $X$ be a random walk in $\mathbb{Z}$. Then, for any age $a > 0$ and any distance $d \in \mathbb{Z}$, we have $\Pr\{X(a) = d\} \leq 2 \cdot \exp\left(-\frac{d^2}{32 \cdot a}\right)$.*

Due to lack of space, the proof of the lemma is omitted (it can be found in Appendix D).

We now compute an estimation of $\Pr\{X(a) = d\}$ when $a$ is sufficiently large. We will use the following asymptotic equivalent of the binomial coefficient, that can be derived by application of the Stirling formula. Let $n_i$ and $m_i$ be two sequences of positive integers such that $n_i \to \infty$, $m_i \to \infty$, and $n_i - m_i \to \infty$ when $i$ grows to infinity. Then

$$\binom{n_i}{m_i} \sim \frac{1}{\sqrt{2\pi}} \cdot \sqrt{\frac{n_i}{m_i \cdot (n_i - m_i)}} \cdot \frac{n_i^{n_i}}{m_i^{m_i} \cdot (n_i - m_i)^{n_i - m_i}}. \tag{4}$$

**Lemma 5** *Let $X$ be a random walk in $\mathbb{Z}$. For any $\zeta > 0$, there exists $d_0 > 1$ such that, for any $|d| \geq d_0$ and $a \geq \frac{d^2}{64 \cdot \ln |d|}$, we have*

$$(1-\zeta) \cdot \sqrt{\frac{2}{\pi \cdot a}} \exp\left(-\frac{3d^2}{4a}\right) \leq \Pr\{X(a) = d\} \leq (1+\zeta) \cdot \sqrt{\frac{2}{\pi \cdot a}} \exp\left(-\frac{d^2}{4a}\right).$$

Due to lack of space, the proof of the lemma is omitted (it can be found in Appendix E). We are now ready to prove of the lower bound of Theorem 1. For the sake of simplicity, let us first assume that the dimension of the lattice is 1. In this case, one can apply the results from the previous section directly. For any $a \geq \frac{3}{4}d^2$ we have

$$\exp\left(-\frac{3d^2}{4a}\right) \geq 1/e\,.$$

Therefore, for any $\zeta > 0$, there exists $d_0$ large enough and $a \geq \frac{3}{4}d^2$ such that Lemma 5 yields:

$$\Pr\{X(a) = d\} \geq \frac{1-\zeta}{e}\sqrt{\frac{2}{\pi}}\frac{1}{\sqrt{a}}.$$

Thus :

$$f(d) = \sum_{a \geq 0} \Pr\{X(a) = d\}\pi(a) \geq \frac{1-\zeta}{e}\sqrt{\frac{2}{\pi}} \sum_{a \geq \frac{3}{4}d^2} \frac{1}{a^{3/2} \cdot \ln^{1+\epsilon}(a)}.$$

More generally, in the $k$-dimensional lattice, let us denote the position of the random walk by $X(a) = (X_1(a), \cdots, X_k(a))$. From the setting of M&F, each $X_i$ is an unbiased random walk in dimension 1, and the $X_i$s are mutually independent. We can thus apply all the results from the previous section independently for each coordinate of $\mathbf{d} = (d_1, \ldots, d_k)$. Assuming that

$$|d_i| \geq d_0 \text{ for all } i \in \{1, \ldots, k\},$$

we can apply Lemma 5 to every dimension. We get:

$$a \geq \frac{3}{4}\|\mathbf{d}\|^2 \implies \forall i \in \{1, \ldots, k\},\, \Pr\{X_i(a) = d_i\} \geq \frac{1-\zeta}{e}\sqrt{\frac{2}{\pi}}\frac{1}{\sqrt{a}}\exp\left(-\frac{3d_i^2}{4a}\right)\,.$$

For $a \geq \frac{3}{4}\|\mathbf{d}\|^2$, we have,

$$\frac{3}{4}\frac{d_i^2}{a} \leq \frac{3}{4}\frac{\|\mathbf{d}\|^2}{a} \leq 1 \text{ and thus } \exp\left(-\frac{3d_i^2}{4a}\right) \geq 1/e.$$

As a consequence, by Lemma 5,

$$\Pr\{X(a) = \mathbf{d}\} = \Pr\{X_1(a) = d_1, \ldots, X_k(a) = d_k\} \geq \left(\frac{1-\zeta}{e}\sqrt{\frac{2}{\pi}}\frac{1}{\sqrt{a}}\right)^k$$

hence

$$f(\mathbf{d}) = \sum_{a \geq 0}\Pr\{X(a) = \mathbf{d}\}\pi_A(a) \geq \left(\frac{1-\zeta}{e}\sqrt{\frac{2}{\pi}}\right)^k \sum_{a \geq \frac{3}{4}\|\mathbf{d}\|^2} \frac{c}{a^{1+(k/2)} \cdot \ln^{1+\epsilon}(a)}.$$

The lower bound is then a direct consequence of the following result with $N = \frac{3\|\mathbf{d}\|^2}{4}$ (the proof of this lemma can be found in Appendix F).

**Lemma 6** *For any $\epsilon > 0$, and any $N \geq e^{2(1+\epsilon)}$, we have*

$$\frac{2/(k+1)}{N^{k/2} \ln^{1+\epsilon} N} \leq \sum_{a \geq N} \frac{1}{a^{1+(k/2)} \ln^{1+\epsilon} a} \leq \frac{2/k}{(A-1)^{k/2} \ln^{1+\epsilon}(N-1)} . \quad (5)$$

Due to lack of space, the proof of the upper bound of Theorem 1 is omitted (it can be found in Appendix G).

## 4 Applications

In the previous section, we have shown that the distribution $f$ of the long-range link lengths is provably converging to a distribution that resembles the $k$-harmonic distribution. In this section, we show that greedy routing can be formally analyzed at the stationary state of this distribution. Greedy routing can be formally analyzed for two reasons: (1) The distribution $f$ of the long-range links constructed by M&F can be bounded formally (cf. Theorem 1); (2) The long-range links resulting from M&F are mutually independent. Based on these two facts, we can establish the theorem below (see proof in Appendix H).

**Theorem 2** *In the $k$-dimensional lattice augmented with the long-range links at the stationary distribution of the dynamical process M&F, the expected number of steps of greedy routing from any source node $s$ to any target node $t$ at distance $d$ is $O(\ln^{2+\epsilon} d)$.*

In the rest of the section, we discuss how M&F can find practical applications to the design of spatial gossip and resource location protocols.

Gossip-based protocols, a.k.a., epidemic algorithms [10], have been introduced as a methodology for designing robust and scalable communication schemes in distributed systems. Roughly, in each step, each node $u$ chooses some other node $v$, and sends a message to it. By applying such scheme at each node, an information originated at some source $s$ will eventually reach its target(s). This methodology can be adapted to various problems, including information spreading, resource location, etc. In [22], Kempe et al. introduced *spatial gossip*, which allowed them to derive efficient solutions for many communication problems. In spatial gossip, nodes are arranged with uniform density in the $k$-dimensional Euclidean space, and, at each step of the gossip protocol, node $u$ chooses node $v$ with probability $\propto 1/d^{\varrho k}$ where $\varrho > 0$ is a fixed parameter, and $d$ is the distance between $u$ and $v$. In particular, it is shown that, for $\varrho \in (1, 2)$, spatial gossip enables to propagate information at distance $d$ in time polylogarithmic in $d$. In [23], Kempe and Kleinberg showed that spatial gossip enables to solve larger classes of problems, including MST construction and permutation routing. In particular, they prove that permutation routing using spatial gossip with $\varrho = 1$ performs in polylogarithmic expected number of steps.

We sketch how the M&F process could facilitate the implementation of the protocols in [22, 23] for networks that take advantage of node mobility, as in, e.g., [7, 20, 21]. For instance, assume a network composed of a set $X$ of fixed nodes and a set $Y$ of moving nodes. Every $x \in X$ connects to the, say, $k_X$ closer neighbors in $X$, and to the, say, $k_Y$ nodes $y \in Y$ that are currently the closest to $x$. Node $x$ keeps all these $y$'s as temporary neighbors, and it regularly checks whether these connections must be preserved. For that purpose, node $x$ regularly flips biased coins (one for each neighbor $y$), and decides whether it should keep a neighbor or not according to the result of this trial. (The bias of the coin is a function $\phi$ of the age of the connection). If $x$ decides to forget some $y$, then $x$ simply replaces $y$ by the node $y' \in Y$ that is currently the closest to $x$. An so on. Assuming that the moving nodes perform random walks, and that all nodes are arranged with uniform density in the $k$-dimensional Euclidean space, Theorem 1 insures that the distances between a node and its moving neighbors are roughly distributed according to a $k$-harmonic distribution. Hence, every fixed node can mimic spatial gossip for $\varrho = 1$ by choosing u.a.r. one if its moving neighbors at each step.

Measuring the impact of the parameters $k_X$ and $k_Y$ on the performances of spatial gossip for $\varrho = 1$, as well as setting up a forgetting function $\phi$ enabling to implement spatial gossip protocols for $\varrho \neq 1$ are beyond the scope of this paper, but are currently under our investigation.

# APPENDIX

## A  Proof of Lemma 1

For all $j \geq 0$, we have:

$$\Pr\{A(t+1) = j \mid A(t) = i\} = \begin{cases} 1 - \phi(j) & \text{if } j = i+1 \\ \phi(i+1) & \text{if } j = 0 \\ 0 & \text{otherwise.} \end{cases}$$

The Markov chain $(A(t))_{t \geq 0}$ is irreducible because any state $i \geq 0$ can be reached from any state $j \geq 0$. Also, $A$ is clearly aperiodic. Let us define the function $\pi$ as follows. For any $a \geq 0$,

$$\pi(a) = \frac{\Pi_{i=1}^{a}(1 - \phi(i))}{\sum_{j \geq 0} \Pi_{i=1}^{j}(1 - \phi(i))}$$

with the convention that the product $\Pi_{i=1}^{0}(1 - \phi(i))$ equals 1. The function $\pi$ is well defined for all $a \geq 0$ by hypothesis on $\phi$. Clearly, $\sum_{a \geq 0} \pi(a) = 1$. We now check that $\pi$ is a stationary distribution. For all $a > 0$, we have

$$\sum_{i \geq 0} \pi(i) \Pr\{A(t+1) = a \mid A(t) = i\} \;=\; \pi(a-1) \cdot (1 - \phi(a-1)) \;=\; \pi(a),$$

and

$$\sum_{i \geq 0} \pi(i) \Pr\{A(t+1) = 0 \mid A(t) = i\} \;=\; \pi(0)\phi(1) + \pi(0) \sum_{i \geq 1} \phi(i+1)\, \Pi_{j=1}^{i}(1 - \phi(j)).$$

Let $B(i) = \Pi_{j=1}^{i}(1 - \phi(j))$ for $i > 0$. We have $1 - \phi(i+1) = B(i+1)/B(i)$, therefore $\phi(i+1)B(i) = B(i) - B(i+1)$. Hence we get

$$\sum_{i \geq 0} \pi(i) \Pr\{A(t+1) = 0 \mid A(t) = i\} \;=\; \pi(0)\Big(\phi(1) + \sum_{i \geq 1}(B(i) - B(i+1))\Big)$$
$$= \pi(0)\,(\phi(1) - B(1))$$
$$= \pi(0).$$

Therefore, $\pi$ is a stationary distribution for $A$, and, since $A$ is irreducible and aperiodic, it is unique. Therefore, $A$ is recurrent positive (see Theorem 3.1, p. 104 in [6]).

## B  Proof of Lemma 2

Let $B(j) = \Pi_{i=1}^{j}(1 - \phi(i))$. We have $B(j) = 1$ for $j = 0, 1, 2$, and

$$B(j) = \frac{2 \ln^{1+\epsilon} 2}{j \ln^{1+\epsilon} j}$$

for $j \geq 3$. Therefore, the series of general term $B(j)$ is finite since $\epsilon > 0$, and $\phi$ satisfies the conditions of Lemma 1. Precisely, we have

$$\sum_{j \geq 0} B(j) = 3 + \sum_{j \geq 3} \frac{2 \ln^{1+\epsilon} 2}{j \ln^{1+\epsilon} j} \leq 3 + \frac{2 \ln 2}{\epsilon} < \infty.$$

Since $\pi(a) = B(a)/\sum_{j \geq 0} B(j)$, the result follows.

## C  Proof of Lemma 3

For any time $t \geq 0$ and any $\mathbf{d}$, we have:

$$\begin{aligned}
\Pr\{C(t) = \mathbf{d}\} &= \Pr\{X(A(t)) = \mathbf{d}\} \\
&= \sum_{a \geq 0} \Pr\{X(a) = \mathbf{d} \text{ and } A(t) = a\} \\
&= \sum_{a \geq 0} \Pr\{X(a) = \mathbf{d}\} \Pr\{A(t) = a\},
\end{aligned}$$

since the Markov chain $A$ is independent of the position of the token. Moreover, since $A$ is recurrent positive (Lemma 1), $A(t)$ converges in variation to $\pi$ when $t$ grows to infinity, that is: $\sum_{a \geq 0} |\Pr\{A(t) = a\} - \pi(a)|$ tends to 0 as $t$ grows to infinity (cf. Theorem 2.1, p. 130 in [6]). Therefore, $\Pr\{A(t) = a\}$ can be replaced by $\pi(a)$ in the above equality when $t$ grows to infinity. Finally $\Pr\{C(t) = \mathbf{d}\}$ is independent of $t$ and its stationary distribution is $f(\mathbf{d})$.

## D  Proof of Lemma 4

First, note that the result is straightforward if $a < |d|$ since the random walk cannot be at distance $d$ in less than $|d|$ time steps. Thus we can assume $a \geq |d|$ in the rest of the proof. Similarly, we can assume $d \neq 0$ since the lemma trivially holds for $d = 0$. Let $\{Y_i, i \geq 1\}$ be a collection of i.i.d. Bernouilli variables that take value 1 with probability $1/2$. Let $T$ be defined by

$$T(a) = Y_1 + \cdots + Y_a. \qquad (6)$$

Thus we get that $X(a)$ and $2T(a) - a$ have the same distribution. Now, $\mathbb{E}[X(a)] = 0$ for any $a \geq 0$. Thus, for any $d \neq 0$,

$$\begin{aligned}
\Pr\{X(a) = d\} &\leq \Pr\{|X(a) - \mathbb{E}[X(a)]| > |d|/2\} \\
&\leq \Pr\{|T(a) - \mathbb{E}[T(a)]| > |d|/4\}.
\end{aligned}$$

The random variable $T(a)$ is the sum of $a$ Bernouilli variables with expectation $1/2$. Thus it has expectation $a/2$, and since $|d|/4$ is less than this expectation, the Chernoff bound of Eq. (3) implies the result.

# E  Proof of Lemma 5

Assume, w.l.o.g., that $d > 0$. Fix $\zeta > 0$. According to the definition of a random walk in $\mathbb{Z}$ we have

$$\Pr\{X(a) = d\} = \frac{1}{2^a}\binom{a}{(a+d)/2}.$$

Let us rewrite $(a+d)/2 = (a/2) \cdot (1+\rho)$ and $a - (a+d)/2 = (a/2) \cdot (1-\rho)$, where $\rho = d/a$. According to Eq. (4) we get that, for any $\zeta' > 0$ with $\zeta' < \zeta$, there exists $d_0$ large enough such that for $|d| \geq d_0$ and $a \geq \frac{d^2}{64 \cdot \ln|d|}$, we have:

$$\Pr\{X(a) = d\} \leq \sqrt{\frac{2}{\pi \cdot a}} \frac{(1+\zeta')}{\sqrt{(1-\rho^2)}} \left(\frac{1}{(1+\rho)^{(1+\rho)} \cdot (1-\rho)^{(1-\rho)}}\right)^{\frac{a}{2}}. \quad (7)$$

On the other hand, for any $x \in (-1, 1)$ we have

$$((1+x)^{(1+x)} \cdot (1-x)^{(1-x)})^{-1} = \exp\left(-(1+x)\ln(1+x) - (1-x)\ln(1-x)\right).$$

As $x$ approaches zero we have $(1+x)\ln(1+x) = x + \frac{x^2}{2} + o(x^2)$, and thus

$$(1+x)\ln(1+x) + (1-x)\ln(1-x) = x^2 + o(x^2).$$

This latter expression can be rewritten as: for any $\nu > 0$ there exists $\eta > 0$ such that:

$$|x| < \eta \implies \exp\left(-(1+\nu)x^2\right) \leq \frac{1}{(1+x)^{1+x}(1-x)^{1-x}} \leq \exp\left(-(1-\nu)x^2\right).$$

Since $\rho = \frac{d}{a} \leq \frac{64 \ln d}{d}$ becomes arbitrarily close to zero for large values of $d$, one can chose $d_0$ large enough so that Eq. (7) holds if one replaces the value inside the bracket by the above upper bound, with $\nu = 1/2$. Hence we get that, for $d \geq d_0$ and $a \geq \frac{d^2}{64 \ln d}$,

$$\Pr\{X(a) = d\} \leq (1+\zeta')\sqrt{\frac{2}{\pi \cdot a}} \frac{1}{\sqrt{1-\rho^2}} \exp\left(-\rho^2 a/4\right).$$

Once again, since $\rho$ is arbitrarily close to zero for large $d$, we can choose $d_0$ large enough so that $(1+\zeta')/\sqrt{1-\rho^2} \leq 1 + \zeta$. The upper bound in the statement of the lemma follows.

Only equivalent forms have been used to establish the upper bound in the statement of the lemma. Thus we can prove the lower bound by applying exactly the same arguments.

# F  Proof of Lemma 6

Let $g : x \mapsto -1/(x^{k/2} \ln^{1+\epsilon} x)$. The derivative of this function satisfies

$$g'(x) = \frac{k}{2} x^{-k/2-1} (\ln^{-(1+\epsilon)} x) + (1+\epsilon) x^{-k/2-1} (\ln^{-(2+\epsilon)} x)$$

$$= \frac{1}{x^{k/2+1} \ln^{1+\epsilon} x} \left( k/2 + \frac{1+\epsilon}{\ln x} \right).$$

Therefore

$$\frac{k}{2} \frac{1}{x^{k/2} \ln^{1+\epsilon} x} \leq g'(x) \leq \frac{k+1}{2} \frac{1}{x^{k/2} \ln^{1+\epsilon} x} \quad \text{if } x \geq e^{2(1+\epsilon)}.$$

As a consequence,

$$\frac{2}{k+1} g'(x) \leq \frac{1}{x^{k/2} \ln^{1+\epsilon} x} \leq \frac{2}{k} g'(x)$$

and

$$-\frac{2}{k+1} g(x) \leq \int_x^\infty \frac{1}{u^{k/2} \ln^{1+\epsilon} u} du \leq -\frac{2}{k} g(x).$$

Eq. (5) follows directly from this latter inequality.

# G  Proof of the upper bound of Theorem 1

Again, let us first consider the simple case of dimension 1. Let $d > 1$. In this context, whenever $a < d^2/(64 \ln d)$, we get by Lemma 4 that

$$\Pr\{X(a) = d\} \leq 2 \cdot \exp(-2 \ln(d)) \leq \frac{2}{d^2}.$$

More generally, let us denote by $i_0$ the dimension that yields the infinity norm of $\mathbf{d}$ (i.e., such that $|d_{i_0}| = \|\mathbf{d}\|$). By applying Lemma 4, we get that if $a \leq \frac{\|\mathbf{d}\|^2}{64 \ln \|\mathbf{d}\|}$ then

$$\Pr\{X_1(a) = d_1, \cdots, X_k(a) = d_k\} \leq \Pr\{X_{i_0}(a) = d_{i_0}\}$$
$$\leq 2/d_{i_0}^2 = 2/\|\mathbf{d}\|^2.$$

For any $\zeta > 0$, there exists $d_0 > 0$ such that if $d_i \geq d_0$ for all $i = 1, \ldots, k$, then we can apply Lemma 5 separately for each dimension. If $a \geq \|\mathbf{d}\|^2 /(64 \ln \|\mathbf{d}\|)$, then

$$\forall i \in \{1, \ldots, k\}, \ \Pr\{X_i(a) = d_i\} \leq (1+\zeta) \cdot \sqrt{\frac{2}{\pi \cdot a}}.$$

Thus, since, for a fixed $a$, the random variables $X_i(a)$ are mutually independent, we get

$$\Pr\{X_1(a) = d_1, \ldots, X_k(a) = d_k\} \leq \left((1+\zeta) \cdot \sqrt{\frac{2}{\pi}}\right)^k \frac{1}{a^{k/2}}.$$

As a consequence,

$$\begin{aligned}
f(\mathbf{d}) &= \sum_{a < \frac{\|\mathbf{d}\|^2}{(64 \ln \|\mathbf{d}\|)}} \Pr\{X(a) = \mathbf{d}\} \pi(a) + \sum_{a \geq \frac{\|\mathbf{d}\|^2}{(64 \ln \|\mathbf{d}\|)}} \Pr\{X(a) = \mathbf{d}\} \pi(a) \\
&\leq \frac{2}{\|\mathbf{d}\|^2} + \left((1+\zeta)\sqrt{\frac{2}{\pi}}\right)^k \sum_{a \geq \frac{\|\mathbf{d}\|^2}{(64 \ln \|\mathbf{d}\|)}} \frac{c}{a^{1+(k/2)} \ln^{1+\epsilon}(a)}.
\end{aligned}$$

One can then complete the proof by using Eq. (5) with $N = \frac{\|\mathbf{d}\|^2}{64 \ln \|\mathbf{d}\|}$.

## H   Proof of Theorem 2

Let $\mathbf{s} \in \mathbb{Z}^k$ be a source node, and $\mathbf{t} \in \mathbb{Z}^k$ be a target node. Assume that the distance between $\mathbf{s}$ and $\mathbf{t}$ in the lattice $\mathbb{Z}^k$ is $\mathrm{dist}(\mathbf{s}, \mathbf{t}) = d$, where dist denotes the $\ell_1$ distance in $\mathbb{Z}^k$. We compute the expected number of steps greedy routing takes before reducing the distance to the target by a factor 2. Let $\mathbf{u} = (u_1, \ldots, u_k) \in \mathbb{Z}^k$ be the current node reached by greedy routing, and let
$$B = \{\mathbf{v} \in \mathbb{Z}^k : \mathrm{dist}(\mathbf{v}, \mathbf{t}) \leq \mathrm{dist}(\mathbf{u}, \mathbf{t})/2\}.$$

The probability $\Pr(\mathbf{u} \to B)$ that $\mathbf{u}$ has its long-range link pointing to a node in $B$ satisfies
$$\Pr\{\mathbf{u} \to B\} = \sum_{\mathbf{v} \in B} \Pr\{\mathbf{u} \to \mathbf{v}\}.$$

We prove a lower bound on this probability. Let $\delta = \mathrm{dist}(\mathbf{u}, \mathbf{t})$. Let
$$S = \{\mathbf{x} = (x_1, \ldots, x_k), x_i \in \{-1, 0, +1\} \text{ for } i = 1, \ldots, k\}.$$

For $\mathbf{c} \in \mathbb{Z}^k$ and $r \geq 0$, let $B(\mathbf{c}, r)$ denotes the ball of radius $r$ centered at $\mathbf{c}$, that is
$$B(\mathbf{c}, r) = \{\mathbf{v} \in \mathbb{Z}^k : \mathrm{dist}(\mathbf{c}, \mathbf{v}) \leq r\},$$
and, for $\mathbf{x} \in S$, define
$$B_{\mathbf{x}} = B(\mathbf{t} + \frac{2\delta}{6k}\mathbf{x}, \frac{\delta}{6k}).$$

We have $B_{\mathbf{x}} \subseteq B = B(\mathbf{t}, \delta/2)$ for any $\mathbf{x} \in S$. Moreover, one can easily show that there exists $\mathbf{x} \in S$ such that for any $\mathbf{v} = (v_1, \ldots, v_k) \in B_{\mathbf{x}}$ and any $i \in \{1, \ldots, k\}$, we have $|u_i - v_i| \geq \delta/(6k)$.

$$\Pr\{\mathbf{u} \to B\} \geq \sum_{\mathbf{v} \in B_{\mathbf{x}}} \Pr\{\mathbf{u} \to \mathbf{v}\} \geq |B_{\mathbf{x}}| \cdot \min_{\mathbf{v} \in B_{\mathbf{x}}} \Pr\{\mathbf{u} \to \mathbf{v}\}.$$

If $\delta \geq 6kd_0$ then $|u_i - v_i| \geq d_0$ for all $i$, and, by Theorem 1, we get that for any $\mathbf{v} \in B_{\mathbf{x}}$,

$$\Pr\{\mathbf{u} \to \mathbf{v}\} \geq \frac{c}{\|\mathbf{u} - \mathbf{v}\|^k \cdot \ln^{1+\epsilon} \|\mathbf{u} - \mathbf{v}\|}.$$

where $\|\cdot\|$ denotes the $\ell_\infty$ norm. Since $\|\mathbf{u} - \mathbf{v}\| \leq \text{dist}(\mathbf{u}, \mathbf{v})$, we get that

$$\Pr\{\mathbf{u} \to \mathbf{v}\} \geq \frac{c}{\text{dist}(\mathbf{u}, \mathbf{v})^k \cdot \ln^{1+\epsilon} \text{dist}(\mathbf{u}, \mathbf{v})}.$$

Now, for any $\mathbf{v} \in B_{\mathbf{x}}$, we have $\mathbf{v} \in B$ and thus $\text{dist}(\mathbf{u}, \mathbf{v}) \leq 3\delta/2$. Therefore,

$$\Pr\{\mathbf{u} \to \mathbf{v}\} \geq \frac{c}{(\frac{3\delta}{2})^k \ln^{1+\epsilon}(\frac{3\delta}{2})}.$$

Since $|B_{\mathbf{x}}| \geq \Omega\left(\left(\frac{\delta}{k}\right)^k\right)$, we get that

$$\Pr\{\mathbf{u} \to B\} \geq \Pr(\mathbf{u} \to B_{\mathbf{x}}) \geq \Omega\left(\frac{1}{\ln^{1+\epsilon}\delta}\right) \geq \Omega\left(\frac{1}{\ln^{1+\epsilon}d}\right).$$

As a consequence, at every intermediate node $\mathbf{u}$ of greedy routing from $\mathbf{s}$ to $\mathbf{t}$, if $\text{dist}(\mathbf{u}, \mathbf{t}) \geq 6kd_0$ then the probability of halving the distance to the target at the next step is at least $\Omega(\frac{1}{\ln^{1+\epsilon}d})$. Since all the long-range links resulting from M&F are mutually independent, we get that the expected number of steps for halving the distance to the target is $O(\ln^{1+\epsilon}d))$. By linearity of the expectation, we get that the total expected number of steps for routing from $\mathbf{s}$ to a node at distance at most $6kd_0$ from the target $\mathbf{t}$ is at most

$$\sum_{i=\lceil \log_2 6kd_0 \rceil}^{\lceil \log_2 d \rceil} \mathbb{E}[\text{halving the distance } \delta \text{ from } 2^{i+1} \text{ to } 2^i]$$

$$\leq \sum_{i=\lceil \log_2 6kd_0 \rceil}^{\lceil \log_2 d \rceil} O(\ln^{1+\epsilon}(2^{i+1}))$$

$$\leq O(\ln^{2+\epsilon} d).$$

Once at distance less than $6kd_0$ to the target, greedy routing completes in $O(1)$ steps, thus the total expected number of steps of greedy routing from $s$ to $t$ is $O(\ln^{2+\epsilon} d)$.